\documentstyle[aps,floats,epsfig]{revtex}
\def\snp{\sigma^{tot}_{\nu p}}
\begin{document}

\twocolumn
\renewcommand{\topfraction}{1.0}
\twocolumn[\hsize\textwidth\columnwidth\hsize\csname
@twocolumnfalse\endcsname
\title{Extensive air showers with TeV-scale quantum gravity}
\author{Luis Anchordoqui, Haim Goldberg, Thomas McCauley, Thomas Paul, 
Stephen Reucroft, and John Swain} 
\address{Department of Physics, Northeastern University, Boston, MA 02115, USA}

\maketitle

\begin{abstract}
One of the possible consequences of the existence of extra
degrees of freedom beyond the electroweak scale is the increase of 
neutrino-nucleon cross sections ($\sigma_{\nu N}$) beyond Standard Model 
predictions. At ultra-high energies this may allow the existence of 
neutrino-initiated extensive air showers.
In this paper, we examine the most relevant observables of such   
showers. Our analysis indicates that the future Pierre Auger Observatory 
could be
potentially powerful in probing models with large compact dimensions. 
\end{abstract}

\vskip2pc]
%%%%%%%%%%%%%%%%%%%%%%%%%%%%%%%%%%%%%%%%%%%%%%%%%%%%%%%%%%%%%%%

Recently, it has become evident that a promising route to reconcile 
high energy particle physics and gravity is to modify the nature of 
gravitational interactions at distances shorter than a millimeter. 
Such a modification can be most simply achieved by introducing extra 
dimensions 
in the sub-millimeter range \cite{ADD}. In this approach the fundamental 
scale of gravity $M_*$ can be lowered all the way to ${\cal O}$ (TeV), 
and the 
observed Planck scale turns out to be just an effective scale valid for 
energies below 
the mass of Kaluza--Klein (KK) excitations. Clearly, while the gravitational 
force has not been directly measured below the millimeter range, Standard 
Model (SM) interactions have been fairly well investigated at this scale; 
so if 
large extra dimensions really exist, one needs some mechanism to prevent 
SM particles from feeling those extra dimensions. Remarkably, there 
are several 
possibilities to confine SM fields (and even gravity) to a 4 dimensional 
subspace (referred to as a 3-brane) within the $(4+n)$  dimensional 
spacetime \cite{trapping}. The provocative new features of this scenario 
have sparked a flurry of activity to assess its experimental 
validity. A brief resum\'e of current theoretical work devoted to higher 
dimensional models includes topics addressing fundamental issues of 
phenomenology \cite{pheno}, cosmology \cite{cosmology}, 
astrophysics \cite{astro}, and 
gravity \cite{gravity}. Moreover, an intense effort to find signatures 
of extra-dimensions in collider data is currently underway \cite{exp}.

Since 1966, a handful of extensive air showers have been
observed corresponding to what seem to be single particles
carrying over $10^{20}$ eV  \cite{reviews}. This, in itself, is
remarkable, as it is difficult or even impossible to explain how
such energies can be attained by conventional acceleration 
mechanisms \cite{bs}. 
Deepening the mystery, it was pointed out by Greisen, Zatsepin and 
Kuz'min \cite{gzk} (GZK) that extremely high energy
($\gtrsim 10^{20}$ eV) cosmic rays, if nucleons and/or nuclei, would lose
energy rapidly through interactions with the cosmic microwave
background (CMB). This leads to the so-called GZK cutoff, which limits the 
propagation distance of these particles to roughly 50 Mpc.  
The difficulty in constructing nearby astrophysical sources that could
accelerate particles to such high energies led to the belief that 
beyond roughly $10^{20}$ eV, no cosmic rays would be detected.
Adding to the puzzle, the arrival directions of these events are 
distributed widely over the sky, with no plausible optical counterparts     
(such as sources in the Galactic plane or in the Local Supercluster).
Furthermore, the ``super-GZK'' data are consistent with an isotropic 
distribution of sources in sharp contrast to the anisotropic distribution of 
light within 50 Mpc from Earth \cite{s-source}.  In conclusion, the current 
picture is very unclear. Thus, it is reasonable to consider whether new physics 
could be at play.

Of particular interest here, the 
extraordinarily high center-of-mass (c.m.) energies achieved at 
the top of the atmosphere are well above those necessary to 
excite the hypothetical KK modes which would reflect a
change in spacetime dimensionality \cite{gravi-burst}.
Hence, a detailed analysis of extensive cosmic ray 
showers, taking into account this departure from previous 
fundamental particle theory, is worthwhile \cite{monopolo}.    

Interestingly enough, if gravity becomes strong at energies of a few TeV,  
virtual graviton exchange can produce relatively large effects on the
high energy scattering cross section, drastically changing the 
neutrino-nucleon interaction \cite{nussinov}. Neutrinos can 
propagate through the 
CMB essentially uninhibited, breaking the GZK 
barrier \cite{sigletal}. Unfortunately, 
within the SM scenario a neutrino incident vertically on the atmosphere would 
pass through it unihibited as well, never initiating an extensive air shower.  
It was already noted that within the extra dimensional framework, the neutrino 
nucleon cross section can approach typical hadronic values at c.m. energies 
$s \agt 400$ TeV, allowing earlier development of a vertical neutrino 
induced shower \cite{domokos,ralston,domokos2}. 
One may wonder whether the growth of the cross section carries with it 
observable deviations from SM predictions. Consistency with current 
experimental data 
requires \cite{haim-tom}, 
\begin{equation}
\sigma (E) \alt 3 \times 10^{-24} \,\frac{E}{10^{19} \,{\rm eV}} \, {\rm cm}^2,
\end{equation}
and this bound certainly does not challenge the neutrinos acquiring a 
hadronic-scale cross section.

A complete theory of massive KK graviton modes is not yet
available, making it impossible to know the exact cross section at
asymptotic energies. Any air shower analysis would thus depend on
reliable guesswork, supplemented with generally acceptable
theoretical principles such as duality, unitarity, Regge behavior
and parton structure.  A simple Born approximation to the elastic
$\nu$-parton cross section \cite{ralston} (which underlies the
total $\nu$-proton cross section) leads, without modification, to
$\snp\sim s^2.$ Unmodified, this behavior by itself eventually
violates unitarity. This  may be seen either by examining the
partial waves of this amplitude, or by noting the high energy
Regge behavior of an amplitude  with exchange of the graviton
spin-2 Regge pole: with  intercept $\alpha(0)=2$, the elastic
cross section
\begin{equation}
\frac{d\sigma_{el}}{dt}\, \sim\, \frac{|A_R(s,t)|^2}{s^2}\, \sim 
s^{2\alpha(0)-2}\,\sim s^2,
\end{equation} 
whereas
\begin{equation}
\sigma_{tot}\, \sim \frac{{\rm Im}[A_R(0)]}{s}\,\sim s^{\alpha(0)-1}\,\sim s,
\end{equation}
so that eventually $\sigma_{el}>\sigma_{tot}.$ Eikonal
unitarization schemes modify these behaviors: in the case of the
tree amplitudes \cite{nussinov} the resulting (unitarized) cross
section $\snp\sim s,$ whereas for the single Regge pole exchange
amplitude, $\snp\sim \ln^2(s/s_0)$ \cite{michael}. However, the
Regge picture of graviton exchange  is not yet entirely
established: both the (apparently) increasing  dominance assumed
by successive Regge cuts due to multiple Regge pole exchange
\cite{nussinov,muzinich}, as well as the presence of
the zero mass graviton can introduce considerable uncertainty in
the eventual energy behavior of the cross section. Hereafter, we
work within the unitarization framework of Ref. \cite{nussinov}
and adopt as our cross section \cite{guenter}
\begin{equation}
\sigma_{\nu N} \approx \frac{4 \pi s}{M_*^4} \approx 10^{-28} 
\left(\frac{M_*}{{\rm TeV}}\right)^{-4} \,\, \left(\frac{E}{10^{19}\,
{\rm eV}}\right)\,{\rm cm}^2.
\label{cs}
\end{equation}

\begin{figure}
\label{nu1}
\begin{center}
\epsfig{file=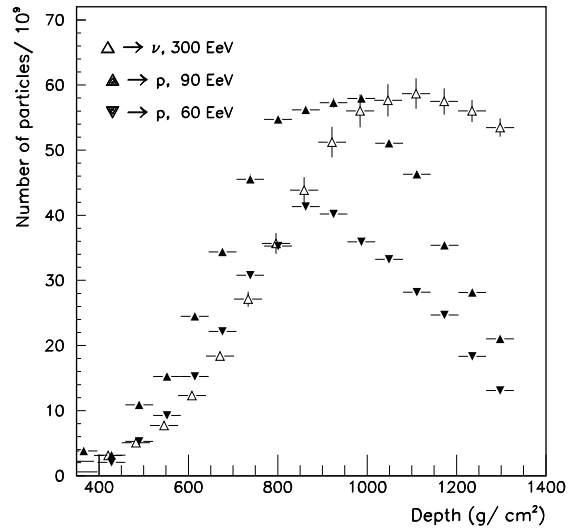,width=8.cm,clip=} 
\caption{Longitudinal development of neutrino and proton showers for different 
primary energies and primary zenith angle 43.9$\circ$. The error bars indicate the standard fluctuations of 
the means.}
\end{center}
\end{figure}

To simulate the consequences of this for $\nu$-induced air showers, 
we  assume that the increase in the cross 
section is driven by the production of minijets \cite{minijets}, 
and we adopt the {\sc sibyll} package to model the fragmentation 
region at ultra high energies \cite{sibyll}. 
In other words, the probability distribution for obtaining $N$ jet pairs
(with $P_T^{{\rm jet}} > P_T^{{\rm min}}$, where $P_T^{{\rm min}}$ is a sharp 
threshold on the transverse momentum above which soft interactions are 
neglected) in a collision at energy $\sqrt{s}$ is computed regarding 
$\nu$-nucleon scattering as a diffractive shadow scattering associated with 
inelastic processes \cite{duranpi}. Particle production comes after the 
fragmentation of hypothetical colorless parton-parton chains mimicking 
that of {\sc sibyll} hadron-hadron scattering. The reader should keep in mind 
the crudeness of this approximation. However, the imposed cutoff on the soft 
processes ensures 
that the inelasticity in any neutrino-nucleon collision is not much larger 
than $y \sim 0.15$ \cite{michael}, justifying the use of the {\sc sibyll} 
package. As we discuss below, most of the expected qualitative 
features in the shower can be quite well reproduced.  The algorithms of 
{\sc aires} (version 2.1.1) \cite{sergio} are slightly modified so as to 
track the particles in the atmosphere. In particular, Eq. (\ref{cs}) is 
translated into the neutrino mean free path 
\begin{equation}
\lambda_\nu = \frac{m_{\rm air}}{\sigma_{\nu {\rm air}}},
\end{equation}
via the standard 8 parameter function used in {\sc aires},
\begin{equation}
\lambda_\nu = P_1 \,\frac{ 1 + P_2\,\, u + P_3 \,\,u^2 + P_4\,\,u^3}{1 
+ P_5 \,\,u 
+ P_6 \,\,u^2 + P_7 \,\, u^3 + P_8 \,\,u^4} \,\, {\rm g}\,{\rm cm}^{-2}.
\end{equation}
Here $m_{\rm air}$ [g] is the mass of an average atom of air, and
$u = \ln E$ [GeV]. The coefficients $P_i$ are listed in Table I 
for different values $M_*$.

Several sets of neutrinos were injected at 100 km above sea level.
The sample was distributed in the energy range of $10^{20}$ eV up 
to $10^{21}$ eV, and was uniformly 
spread in the interval of 0$^{\circ}$ to 60$^{\circ}$ zenith angle at
the top of the atmosphere. All shower particles with energies above the 
following thresholds were
tracked: 750 keV for gammas, 900 keV for electrons and positrons, 10
MeV for muons, 60 MeV for mesons and 120 MeV for nucleons.
The results of these simulations were processed with the help of the 
{\sc aires} analysis package.

\begin{figure}
\label{nu2}
\begin{center}
\epsfig{file=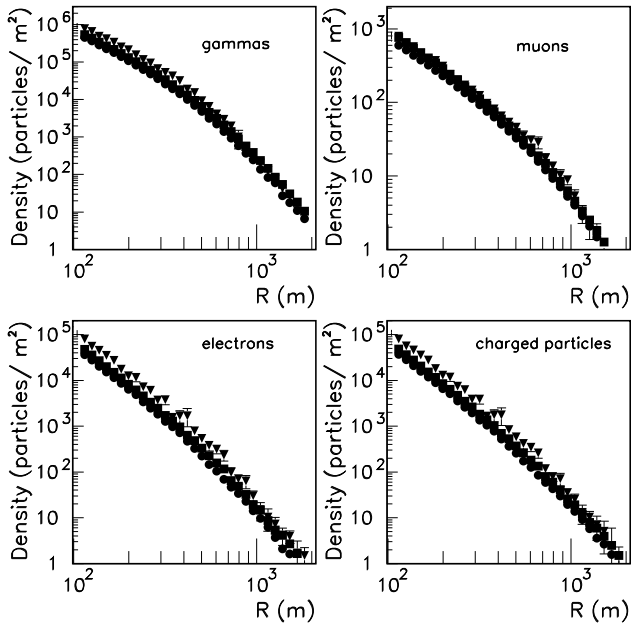,width=9.5cm,clip=} 
\caption{Lateral distributions of vertical 300 EeV neutrino-induced 
showers (triangles), 60 EeV proton-induced showers (circles), and 60 EeV 
iron-induced showers (squares). 
The error bars indicate the RMS fluctuations.}
\end{center}
\end{figure}

Figure 1 shows the total number of charged particles 
versus atmospheric depth averaged over 25 showers for the case of 
a 300~EeV neutrino at $M_*$=1~TeV.  For comparison, proton-induced 
showers at 60 and 90~EeV are shown on the same figure.  As showers 
initiated by neutrinos 
typically start later than proton-induced showers, the longitudinal
development tends to level off after reaching a maximum, in contrast
to a standard air shower which decreases more rapidly after reaching a 
maximum. 
The number of charged particles produced in the cascade depends on 
the amount of energy deposited in the atmosphere by the primary.
Neutrinos at the energy and mass scale 
shown in the figure typically suffer 2 interactions in the atmosphere;
any energy remaining after this is undetected.  By comparing the
neutrino-induced showers to the proton-induced showers shown in the figure, one
can roughly estimate the inelasticity to be $0.1 < y < 0.15$.  This is
consistent with the estimates of reference~\cite{michael}.\footnote{It is 
important to stress that the maximum number of charged particles produced 
in a proton-induced shower does not depend on the hadronic interaction 
model \cite{prdhi}, making the present estimate on the 
inelasticity quite reliable.}

Figure 2 shows the lateral distributions for vertical showers produced
by 300~EeV neutrinos, 60~EeV protons, and iron nuclei of 60~EeV. 
At 50~m from the core, 
the ratio of the number of charged particles in the neutrino shower
to that in the proton shower is $\approx 2$, whereas it is 
$\approx 1.5$ in $\nu$/$^{56}$Fe showers. At about 1~km 
from the core these ratios reduce to $\approx 1.1$ and $\approx 0.7$, 
respectively. This is significant 
since experiments
which rely on surface detectors to determine shower parameters typically 
use samples taken on the order of 1~km from the core, and thus would
not be able to easily distinguish between these particle species.

\begin{figure}
\label{1}
\begin{center}
\epsfig{file=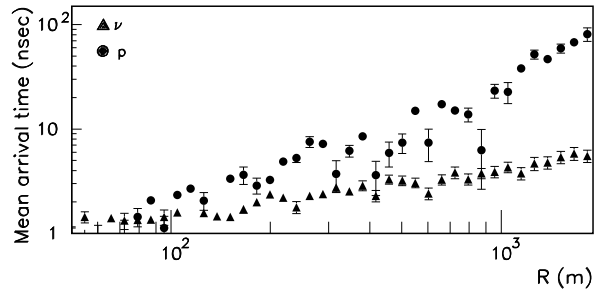,width=8.cm,clip=} 
\caption{Arrival times for charged particles in 
vertical 300~EeV neutrino and 60~EeV proton showers normalized at 50~m 
from the shower core. The error bars indicate the RMS fluctuations.}
\end{center}
\end{figure}

Figure 3 shows the radial dependence of the mean arrival time of muons  
for  showers initiated by 300~EeV neutrinos and 60~EeV protons.
It can be readily seen from the comparison that the proton-induced showers 
exhibit 
larger fluctuations than the neutrino-induced showers. Besides, 
each profile presents a well defined slope that characterizes the shower front
and comprises a  signature of the primary species. In particular, a neutrino 
interacts in the atmosphere only once or twice, and consequently the muons 
reach the ground with a relatively short time delay. 

The simulated neutrino showers discussed so far deposit far less
energy in the atmosphere than the most energetic of the observed 
cosmic ray events.  A natural question is then what the 
shower profile would look like for a neutrino whose energy and mean 
free path are such that it would deposit roughly the same energy 
as observed in the highest energy event~\cite{fe}.  

At this stage, it is important to point out that within the SM framework 
neutrinos are produced at extremely high energies, 
typically by the weak decay of pions or other hadrons. Thus, one needs 
protons to be accelerated to energies a few orders of magnitude even higher. 
In scenarii involving {\it precocious unification} \cite{zk}, 
there may be alternatives to decay chains for producing super-GZK 
neutrinos at the source.

Figure 4 shows the longitudinal development of a 900~EeV
neutrino-induced shower with a fundamental mass scale $M_* = 1.3$~TeV.  
We stress that such a scale is above the lower bound for $M_*$ derived from the 
expected flux of neutrinos and current non-observation of horizontal 
air showers \cite{guenter}. The total energy
deposited in the atmosphere (after 2 interactions) is of the
same order as the Fly's Eye event, but the shower maximum occurs, as
expected, significantly later. 

\begin{figure}
\label{nu4}
\begin{center}
\epsfig{file=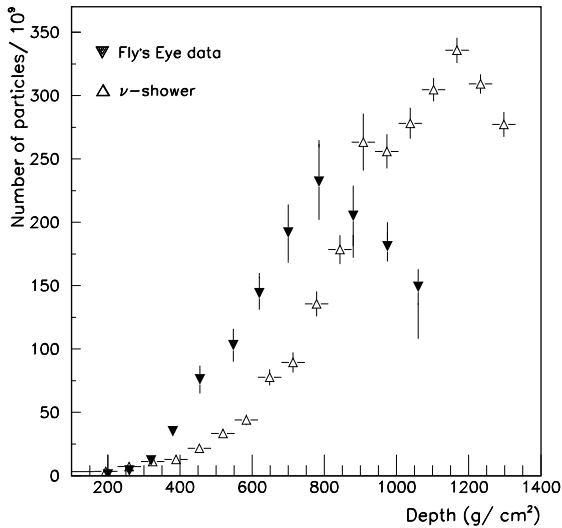,width=8.cm,clip=} 
\caption{The longitudinal development of a 900 EeV neutrino-induced shower 
is shown together with the experimental data reported by Fly's Eye.   
The error bars in the simulated points indicate the standard fluctuations of 
the means.}
\end{center}
\end{figure}

In summary, it has been proposed~\cite{domokos,ralston,domokos2} 
that the GZK cutoff can be skirted if the progenitors of the most energetic 
air showers are neutrinos.  Under this hypothesis, the neutrino-nucleon
cross section is increased by the presence of extra dimensions,
allowing the neutrinos to interact in the atmosphere.  Simulations indicate 
that neutrino-induced showers at energies 
of a few hundred EeV would exhibit signatures distinct from 
those of proton (or nucleus) induced showers that deposit a similar 
amount of energy in the atmosphere.  Similarly, if there are neutrinos 
energetic enough to deposit as much energy in the atmosphere as
is observed in the highest energy events, it appears they too
may have unique signatures.  In fact, any physics beyond
the standard model that increases the neutrino-nucleon cross
section should affect shower observables
like longitudinal profile (measured with fluorescence detectors)
and ground particle distributions (measured with surface detectors).
This article contains some qualitative discussion of relevant  
observables of neutrino-induced showers.  As far as we are aware, no
showers have been observed which are consistent with these features.
If candidates are eventually discovered, of course it will be necessary to 
carry out a much more detailed simulation than the one presented here.
We note that future hybrid detectors such as the Pierre Auger Observatory~\cite{auger} 
will be in an exceptional position to search for such phenomena.

{\bf Note added:} After this paper was written, it was stressed that 
extremely high energy (300 EeV) neutrinos with larger cross section 
($s^2$ rise) can create showers that would look like the highest energy 
event \cite{jain}. If this is the case, it should also be stressed that 
neutrinos 
of a few tens of EeV could induce vertical air showers with very 
distinctive 
profiles. In Fig. 5 we show the longitudinal development of showers 
initiated by neutrinos of $E = 5 \times 10^{19}$ eV.\footnote{To 
compute the simulation we adopt the cross section growth used in Ref. 
\cite{jain} to reproduce the Fly's Eye data.}
For comparison we also 
show showers induced by gamma rays and protons of $E = 5 \times 10^{18}$ eV.
It is easily seen that within this framework a 50 EeV neutrino 
shower presents its own signature \cite{gamma}.

The question of whether the interaction cross section of neutrinos with matter
could be greatly enhanced (via massive spin-2 exchange) at high energies, 
is yet undecided. Observation of deeply penetrating showers with 
$5 \times 10^{18}$ eV deposited in the atmosphere  would 
give an experimental and definite answer to this question. 
As an immediate spinoff, we have the converse fact, i.e., that if there 
were no possible candidate which could be associated with a neutrino shower, 
then it should be understood as a serious objection to the  hypothesis 
of neutrinos as progenitors of the ``super-GZK'' events. We strongly 
recommend that the Fly's Eye data be re-analyzed searching for evidence of 
neutrino showers.

%%%%%%%%%%%%%%%%%%%%%%%%%%%%%%%%%%%%%%%%%%%%%%%%%%%%%%%%%%%%%%
\acknowledgements{
We would like to thank M\'aximo Ave, Anal\'{\i}a Cillis, Gabor Domokos, 
Michael Kachelrie{\ss}, Zurab Kakushadze, Susan Kovesi-Domokos,
Jeremy Lloyd-Evans, Doug McKay, Michael Plumacher, John Ralston, 
Lisa Randall, Sergio Sciutto, Robert Shrock, and 
Ricardo V\'azquez for useful discussions/correspondence. 
This work was partially supported by CONICET (Argentina) 
and the National Science Foundation.}
%%%%%%%%%%%%%%%%%%%%%%%%%%%%%%%%%%%%%%%%%%%%%%%%%%%%%%%%%%%%%%%

\begin{figure}
\label{nu5}
\begin{center}
\epsfig{file=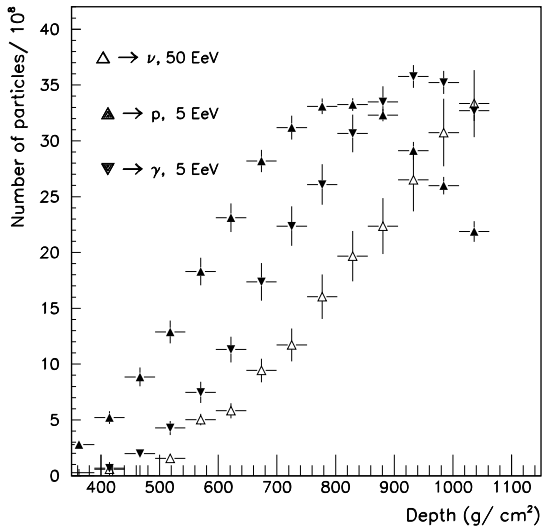,width=8.cm,clip=} 
\caption{Longitudinal development of 50 EeV neutrino-induced 
showers, 5 EeV proton-induced showers, and 5 EeV 
gamma-ray showers. The error bars indicate the standard fluctuations of 
the means.}
\end{center}
\end{figure}

%%%%%%%%%%%%%%%%%%%%%%%%%%%%%%%%%%%%%%%%%%%%%%%%%%%%%%%%%%%%%%%%%%%%

\begin{onecolumn}
\begin{table}
\caption{Coefficients for mean free path parametrization}
\begin{tabular}{ccccccccc}
%\hline\hline 
$M_*$ [TeV] & $P_1$ & $P_2$ & $P_3$ & $P_4$ & $P_5$ & $P_6$ & $P_7$ 
& $P_8$
\\ \hline 
1 & -14657 & -2254.4 & -13.931 & 3.3530 & -1236.7 & -814.89 & -4.6945 & 
1.7814 \\
1.2 & 5654.4 & 1130000 & 1393 & -1417.3 & -1724000 & -124980 & 
100.44 & 316.09 \\
1.3 & 6638.5 & 307640 & 355.94 & -366.14 & -1499700 & -19822 
& 845.46 & 91.015\\
%\hline \hline
\end{tabular}
\end{table}
\end{onecolumn}


\begin{thebibliography}{99}


\bibitem{ADD}  
N. Arkani-Hamed, S. Dimopoulos and G. Dvali, Phys.
Lett. B {\bf 429}, 263 (1998); I. Antoniadis, N. Arkani-Hamed,
S. Dimopoulos and G. Dvali, Phys. Lett. B {\bf 436}, 257 (1998).

\bibitem{trapping} 
G. Dvali and M. Shifman, Phys. Lett. B {\bf 396}, 64 (1997), 
{\it erratum ibid} {\bf 407}, 452 (1997);
L. Randall and R. Sundrum, Phys. Rev. Lett. {\bf 83}, 
4690 (1999); B. Bajc, G. Gabadadze, Phys. Lett. B {\bf 474}, 282 (2000);
G. Dvali, G. Gabadadze, [hep-th/0008054].


\bibitem{pheno} See for instance, 
G. F. Giudice, R. Rattazzi 
and J. D. Wells, Nucl. Phys. B {\bf 544}, 3 (1999); T. Han, J. 
D. Lykken and R. J. Zhang, Phys. Rev. D {\bf 59}, 105006 (1999); 
J. L. Hewett, Phys. Rev. Lett. {\bf 82} 4765 (1999); E. A. Mirabelli, M. 
Perelstein and M. E. Peskin, Phys. Rev. Lett. {\bf 82}, 2236 (1999); 
T. G. Rizzo, [hep-ph/9910255]; S. Cullen, M. Perelstein and M. E. Peskin, 
Phys. Rev. D {\bf 62}, 055012 (2000); L. Randall and R. Sundrum,
 Phys. Rev. Lett. {\bf 83}, 3370 (1999); J. Lykken and 
L. Randall, JHEP {\bf 0006}, 014 (2000).   


\bibitem{cosmology} See for instance, C. Cs\'aki, M. Graesser, 
C. Kolda, J. Terning, Phys. Lett. B {\bf 462}, 34 (1999);
J. M. Cline, C. Grojean and G. Servant, Phys. Rev. Lett. {\bf 83},
4245 (1999); C. Cs\'aki, M. Graesser, 
L. Randall and J. Terning, Phys. Rev. D {\bf 62}, 
045015 (2000); S. Nojiri and S. Odintsov, JHEP {\bf 0007}, 049 (2000); 
L. A. Anchordoqui, C. Nu\~nez, K. Olsen, JHEP {\bf 0010}, 050 (2000) 
[hep-th/0007064]; S. W. Hawking, T. Hertog and H. S. Reall, [hep-ph/0010232]. 

\bibitem{astro} See for instance, S. Cullen and 
M. Perelstein, Phys. Rev. Lett. 
{\bf 83}, 268 (1999); N. Arkani-Hamed, S. Dimopoulos, G. Dvali and N. Kaloper,
[hep-ph/9911386];  V. Barger, T. Han, C. Kao, R. J. Zhang, Phys. Lett. 
B {\bf 461}, 34 (1999); G. C. McLaughlin, Phys. Lett. B {\bf 470}, 157 (1999);
S. Cassisi, V. Castellani, S. Degl'Innocenti, G. Fiorentini, B. Ricci, Phys. 
Lett. B {\bf 481}, 323 (2000).



\bibitem{gravity} See for instance, W. D. Goldberger and M. B. Wise, 
Phys. Rev. Lett. {\bf 83}, 4922 (1999)
A. Chamblin, S. W. Hawking and 
H. S. Reall, Phys. Rev D {\bf 61}, 065007 (2000);  J. Garriga and 
T. Tanaka, Phys. Rev. Lett. {\bf 84} 2778 (2000); S. Nojiri, S. D. Odintsov 
and S. Zerbini, Phys. Rev D {\bf 62}, 064006 (2000); S. B. Giddings, 
E. Katz and L. Randall, JHEP {\bf 0003}, 023 (2000); S. W. Hawking, 
T. Hertog and H. S. Reall, Phys. Rev. D {\bf 62}, 043501 (2000);
D. Langlois, R. Maartens and D. Wands, Phys. Lett. B {\bf 489}, 259 (2000); 
C. D. Hoyle et al., [hep-ph/0011014]. 

\bibitem{exp} M. Acciarri et al. (L3 Collaboration), Phys. Lett. B {\bf 470},
281 (1999) [hep-ex/9910056]; C. Adloff et al. (H1 Collaboration),
[hep-exp/0003002]; B. Abbott et al. (D\O $\,\,$ Collaboration), 
[hep-ex/0008065].



\bibitem{reviews} S. Yoshida and H. Dai, J. Phys {\bf G24} 905
(1998); M. Nagano, A. A. Watson, Rev. Mod. Phys. {\bf 72}, 659 (2000).

\bibitem{bs} For a comprehensive review on the origin of the highest energy 
cosmic rays the reader is referred to P. Bhattacharjee and G. Sigl, Phys. Rep. 
{\bf 327} 109 (2000).


\bibitem{gzk} K. Greisen , Phys Rev. Lett. {\bf 16} 748 (1966); 
G.T. Zatsepin and V.A. Kuz'min, Pis'ma Zh. Eksp. Teor.
Fiz. {\bf 4} 114 (1966) [JETP Lett. {\bf 4} 78 (1966)].

\bibitem{s-source} The observed near-isotropy of the distribution can be 
explained, within a single-source hypothesis, postulating a 
galactic wind  or a large extragalactic magnetic field. 
E. -J. Ahn, G. Medina-Tanco, P. L. Biermann and T. Stanev, 
[astro-ph/9911123]; G. Farrar and T. Piran, [astro-ph/0010370].
None of these models, however, could explain directional 
clustering as discussed by N. Hayashida et al. (AGASA Collaboration), Phys. 
Rev. Lett. {\bf 77}, 1000 (1996); M. Takeda et al., Astrophys. J. {\bf 522},
225 (1999) [astro-ph/9902239 and astro-ph/0008102]; Y. Uchihori et al., 
Astropart. Phys. {\bf 13}, 151 (2000). For a recent analytic analysis, see 
H. Goldberg and T. J. Weiler, [astro-ph/0009378]. 
 

\bibitem{gravi-burst} In considering the 
exchange of gravitons (KK modes), one should distinguish the following 
two scenarios: (i) In the canonical example of \cite{ADD}, 
KK gravitons are couple extremely weakly, 
and the observational effects arise because of the very large 
multiplicity of states due to their 
fine splittings. (ii) In the Anti--de Sitter bulk scenario 
(see Randall-Sundrum in Ref. \cite{pheno}), each exited state coupling 
is ${\cal O} (E/{\rm TeV})$, and thus single KK modes could be detected 
via their decay products. Future cosmic ray data could play an important 
role in testing the latter. H. Davoudiasl, J. L. Hewett and T. G. Rizzo, 
[hep-ph/0010066].


\bibitem{monopolo} The influence of the extra-dimensional scenario 
on  monopole induced showers was reported elsewhere. 
L. A. Anchordoqui, T. P. McCauley, S. Reucroft and J. Swain, 
Phys. Rev. D {\bf 63}, 027303 (2001) [hep-ph/0009319].

\bibitem{nussinov}
S. Nussinov and R. Shrock, Phys. Rev. D {\bf 59} 105002 
(1999). 


\bibitem{sigletal} A remarkable correlation between the arrival direction
of cosmic rays above $10^{20}$ eV and high redshift compact radio quasars 
seems to support the neutrino hypothesis. Such a correlation, however, 
diminishes when considering only the highest energy events 
($E>8 \times 10^{19}$ eV at 1-standard deviation) that have no contamination 
from the expected proton pile-up around the photopion production threshold. 
G. R. Farrar and P. Biermann, Phys. Rev. Lett. {\bf 81}, 3579 (1998); 
G. Sigl et al., [astro-ph/0008363]; A. Virmani et al., [astro-ph/0010235].


\bibitem{domokos}  G. Domokos and S. Kovesi-Domokos, Phys. Rev. 
Lett. {\bf 82}, 1366 (1999).

\bibitem{ralston} P. Jain, D. W. McKay, S. Panda, and J. P. Ralston, 
Phys. Lett. B {\bf 484}, 267 (2000).

\bibitem{domokos2} G. Domokos, S. Kovesi-Domokos and P. T. 
Mikulski [hep-ph/0006328]. 

\bibitem{haim-tom} H. Goldberg, T. J. Weiler, Phys. Rev. D {\bf 59}, 
113005 (1999).

\bibitem{michael} 
M. Kachelrie{\ss} and M. Plumacher, Phys. Rev. D {\bf 62}, 103006 (2000) 
[astro-ph/0005309].


\bibitem{muzinich} I. J. Muzinich and M. Soldate, Phys. Rev. D {\bf 37},
359, (1988).

\bibitem{guenter} C. Taylor, A. Olinto and G. Sigl, Phys. Rev. D {\bf 63},
055001 (2001) [hep-ph/0002257]. 


\bibitem{minijets} T. K. Gaisser and T. Stanev, Phys. Lett. 
B {\bf 219}, 375 (1989).

\bibitem{sibyll} R. S. Fletcher, T. K. Gaisser, P. Lipari and T.
Stanev, Phys. Rev. D {\bf 50}, 5710 (1994).


\bibitem{duranpi} L. Durand and H. Pi, Phys. Rev. Lett. {\bf 58}, 303 (1987).


\bibitem{sergio} S. J. Sciutto, in {\it
Proc. XXVI International Cosmic Ray Conference}, (Edts. D. Kieda, M. Salamon,
and B. Dingus, Salt Lake City, Utah, 1999) vol.1, p.411, [astro-ph/9905185]. 




\bibitem{prdhi} L. A. Anchordoqui, M. T. Dova, L. N. Epele, S. J. Sciutto,
Phys. Rev. D {\bf 59}, 094003 (1999). See in particular Fig. 8.

\bibitem{fe} D. J. Bird et al., Astrophys. J. {\bf 441}, 144 (1995).

\bibitem{zk} Z. Kakushadze, Nucl. Phys. B {\bf 548}, 205 (1999); 
{\bf 552}, 3 (1999); {\bf 551}, 549 (1999). See also,
\cite{domokos}.

\bibitem{auger} For an overview of the Auger project see for instance,
D. Zavrtanik, Nucl. Phys. B Proc. Suppl. {\bf 85}, 324 (2000).


\bibitem{jain} A. Jain, P. Jain, D. W. McKay, and J. 
P. Ralston, [hep-ph/0011310].

\bibitem{gamma} It is worthwhile to remark that at these 
energies the CMB is completely opaque to the propagation of gamma rays. 
See for instance, 
R. J. Protheroe and P. A. Johnson, Astropart. Phys. {\bf 4}, 
253 (1996).


\end{thebibliography}
\end{document}